\documentclass[useAMS,usenatbib]{mn2e}
\usepackage{amsfonts,amsmath,amssymb,bm,epsfig,url}
\usepackage[dvips]{color}
\usepackage{graphicx}
\usepackage{dcolumn}
\usepackage[pagebackref=false, colorlinks=true]{hyperref}
\usepackage{textcomp}
\usepackage{float}
\usepackage{subfigure}

\definecolor{redish}{rgb}{0.7,0.2,0.0}  
\definecolor{bluish}{rgb}{0.2,0.5,0.8}

\hypersetup{linkcolor=redish,          
                  citecolor=blue,        
                  filecolor=magenta,      
                  urlcolor=bluish}   
                  
\def \f{\frac}
\def \o{\omega}

\def \a{\alpha}
\def \b{\beta}
\def \O{\Omega}
\def \g{\gamma}

\def \p{\partial}

\def \S{\Sigma}

\title[A tilted and warped accretion disc around a black hole]
{A tilted and warped inner accretion disc around a spinning black hole: an analytical solution}
\author[C. Chakraborty and S. Bhattacharyya]
{Chandrachur Chakraborty\thanks{chandrachur.chakraborty@tifr.res.in (CC)} 
and  Sudip Bhattacharyya\thanks{sudip@tifr.res.in (SB)}\\
Tata Institute of Fundamental Research, Mumbai 400005, India}
\begin{document}

\date{}

\pagerange{\pageref{firstpage}--\pageref{lastpage}} \pubyear{2016}

\maketitle

\label{firstpage}

\begin{abstract}
Inner accretion disc around a black hole provides a rare, natural probe to 
understand the fundamental physics of the strong gravity regime. A possible 
tilt of such a disc, with respect to the black hole spin equator, is important.
This is because such a tilt affects the observed spectral 
and timing properties of the disc X-ray emission via Lense-Thirring precession,
which could be used to test the 
theoretical predictions regarding the strong gravity. Here, we analytically 
solve the steady, warped accretion disc equation of Scheurer and Feiler (1996),
and find an expression of the radial profile of the disc tilt angle.
In our exact solution, considering a prograde disc around a slowly spinning 
black hole, we include the inner part of the disc, which was not done earlier in this formalism.
Such a solution is timely, as a tilted inner disc has recently been inferred
from X-ray spectral and timing features of the accreting black hole H1743--322.
Our tilt angle radial profile expression includes observationally measurable parameters, such
as black hole mass and Kerr parameter, and the disc inner edge tilt angle $W_{\rm in}$,
and hence can be ideal to confront observations. Our solution shows that the
disc tilt angle in $10-100$ gravitational radii is a significant fraction of the 
disc outer edge tilt angle, even for $W_{\rm in} = 0$. Moreover, tilt angle radial profiles
have humps in $\sim 10-1000$ gravitational radii for some
sets of parameter values, which should have implications for observed X-ray 
features.
\end{abstract}


\section{Introduction}\label{Introduction}
Local inertial frames are dragged due to the rotation of the spacetime,
which is known as the Lense-Thirring (LT) effect (\cite{lt}.)
In 1975, \citet{bp} suggested that this relativistic LT effect can make important 
changes in the geometrically thin accretion disc structure 
around a spinning Kerr black hole,
if the disc is not in the equatorial plane with respect to the black hole spin axis.
More specifically, the LT effect causes a gradual transition of the inner part
of such a tilted disc into the equatorial plane of the black hole.
This mechanism is known as the Bardeen-Petterson (BP) effect.
The LT effect also twists the disc, while the disc viscosity limits this 
twisting \citep{fmw}. According to the BP effect, the competition between the LT effect and viscous
forces divide the disc into three regions: (1) the outer part which remains tilted
with respect to the black hole spin, (2) the inner part which becomes 
aligned with the black hole spin due to a strong LT effect, and 
(3) a twisted transition region, called a
`warp', between the above two regions. Note that the outer part remains tilted,
because the LT precession rate drops off rapidly with the radial distance ($R$)
from the black hole centre (roughly as $1/R^3$). 

After \citet{bp} proposed the above mentioned effect, 
\citet{p71} made detailed formulation of the time-independent disc equations,
and \citet{p72} applied this formulation to the X-ray binary systems
to study their properties. Then \citet{p78} derived the time-dependent equations of
the warped disc, and studied properties of such a disc. \citet{mp} described
an algorithm to solve for the time-evolution of twisted disc equations. 
Using this, and assuming that 
the X-ray intensity variation of the X-ray binary Her X--1 
with a 35 day cycle is due to the periodic obscuration of a precessing disc,
these authors found constraints on the viscosity in the disc.
Theoretically they also predicted 
a range of the viscous anisotropy parameter $\a_a$\footnote{$\a_a$ is the viscous anisotropy
defined as the ratio of the `vertical' to the `horizontal' coefficients of viscosity.},
which is $\f{1}{3}\lesssim \a_a \lesssim 3$.

Later, \citet{hbs} pointed out several {\it mathematical} inconsistencies in
the derivations of \citet{bp}, \citet{p71} and \citet{p78}, and
modified the pair of ``twist equations'' of the above mentioned 
three papers in a single twist equation using a complex quantity.
This complex quantity $W$ includes both the tilt angle ($\beta$) and
the twist angle ($\gamma$) in a compact manner, as $W = \beta e^{i \gamma}$ ($i=\sqrt{-1}$).
But \citet{pp} pointed out that
angular momentum did not appear to be conserved globally in the \citet{hbs} formulation.

After this, \citet{p92} successfully introduced a simple set of twisted disc equations
in which the time evolution was governed by two viscosities (mentioned 
in the footnote). In the latter part of this paper, LT effect was introduced 
and numerical simulations were given to study the 
alignment timescale of the disc due to the BP effect.
Later, \citet{sf} (hereafter SF96) 
re-wrote Pringle's (1992) equation with LT effect removing the time-dependent term
to present the steady state solution of a warped accretion disc.
The prescription of SF96 successfully considered the necessary physical effects,
and several researchers have used their formalism with various assumptions
to study the solution of a warped disc. 
For example, the disc viscosity was
taken as a constant in SF96, but \citet{chen} have recently modified the solution
taking the viscosity as a function of $R$. This modification in viscosity has
provided a new solution, but without any new physical behaviour of a warped disc.
However, both the above mentioned solutions \citep{sf,chen} are for $R$
values much larger than the disc inner edge radius.

There have been other lines of investigations on tilted discs as well.
For example, \citet{zhul} \citep[see also, ][]{zhu}, from their 
GRMHD simulations and semi-analytic models, found no evidence of BP alignment
for a prograde, tilted disc, but found an evidence of partial alignment
for a retrograde disc. Besides, \citet{nx} suggested that a disc could be
broken into a set of distinct planes with only tenuous flows connecting them,
if the LT torque is higher than the viscous torque.

Although, \citet{bp} predicted that the inner accretion disc
should be in the equatorial plane with respect to the black hole spin axis,
a recent X-ray observation has suggested a tilted disc \citep{Ingrametal2016}.
Before discussing this, let us provide a brief background on two relevant
X-ray features. A prominent spectral feature, which is believed to originate 
when hard X-ray photons are reflected from the inner part of the
geometrically thin accretion disc around a black hole, is an Fe K$\alpha$ broad emission
line near $6.4$ keV. Such a line
is broadened and skewed by the Doppler and relativistic effects, and hence is a useful
tool to measure the black hole spin (\citet{Miller2007} and references therein).
The quasi-periodic oscillations (QPOs) of
the X-ray intensity in a frequency range spanning several orders of magnitude are also
observed from accreting black holes. At least some of these timing features are believed
to originate from various natural disc oscillations, including the LT precession, from
the inner part of the accretion disc (\citet{BelloniStella2014} and references therein).
\citet{Ingrametal2016} has recently reported that 
the broad relativistic Fe line energy from the accreting black hole H1743--322
systematically varies with the phase of a QPO. This could be explained assuming 
the LT precession of a tilted thin accretion disc, which can act as the reflector to give
rise to the observed Fe line \citep[e.g., see sections 6.1 and 6.3 of ][]{Ingrametal2017}.
Therefore, the discovery of \citet{Ingrametal2016} suggests a tilted inner disc.

On the theoretical side, we also note that the equation (9) of \citet{bp}
\citep[or its modified form, i.e., equation (11) of ][]{p73},
being a set of second order differential equations, would give
two solutions of the tilt angle $\beta$. In one case $\beta$ increases with $R$ and in another case
$\beta$ decreases with $R$, as clearly written after equation (11) of \citet{p73}.
But these authors chose only one of these solutions, the one with $\beta$ increasing with $R$,
and such a solution was not shown to be unique.
Note that the solutions do not have to be regular at the origin \citep[as demanded by ][]{bp},
because a disc cannot extend inside the innermost stable circular orbit (ISCO).
Therefore, \citet{bp} did not rule out the possibility of a tilt of the inner disc. Similarly,
\citet{p73} ensured an aligned inner disc solution simply by choosing suitable boundary 
conditions \citep[see after equation (11) of ][]{p73}. In support of this choice, 
this paper proposed that a misaligned twisted inner disc resulted from the other
(discarded) solution would be unstable due to the high X-ray flux near the compact object.
But observational evidence is required to substantiate this guess, particularly for black holes
which cannot irradiate the inner disc from a hard surface like a neutron star.

Motivated by the observational indication of a tilt of the inner disc, as well as 
the theoretical aspects discussed above, here we analytically study the accretion
disc properties by keeping the tilt angle at the disc inner edge a free parameter. This could
be useful to compare our results with observations, when an inner disc tilt angle
with respect to the black hole spin equator is observationally inferred \citep[for example,
in section 6.3 of ][]{Ingrametal2017}.
We explore the steady state solution of a tilted and twisted
accretion disc. Our prescription is based on the Pringle's (1992) equation
(but not his formalism for solution, ways and assumptions which had been made in
section 3 of his paper),
removing the time-dependent part but adding the LT precession as mentioned 
in SF96.
While our solution is valid for the entire disc, we particularly focus on
the inner disc. Note that both SF96 and \cite{chen} dropped from their equations 
a certain term $C_1$, which is connected with the condition of the inner disc.
While this made the solutions of their equations much simpler, this also made
those solutions unsuitable for the inner disc. We, on the other hand, 
find an analytical steady state solution of a tilted disc, without dropping
$C_1$ (but making the solution linear in the Kerr parameter)
and this does not violate any
basic principle which was used by Pringle (1992) and SF96.
Hence, our solution is suitable for the inner disc.

The paper is organized as follows. In Section~\ref{c}, we analytically solve an
equation of a tilted/warped disc. 
We present and discuss the tilt angle radial profiles in Section~\ref{results}.
Finally, we summarize this work and briefly describe the implications of our finding
in Section~\ref{conclu}.

\section{Tilted and warped disc solution}\label{c}

\subsection{The tilted/warped disc equation by Scheurer and Feiler (1996)}\label{ca}

We assume that the viscosity $\nu$ is independent of the radial distance $R$, and the angle 
between the angular momentum vectors of the black hole and accretion disc
is small at large $R$. These two basic assumptions were also used in SF96.
Following Pringle's (1992) equation with Lense-Thirring precession, one can
rewrite the basic tilted/warped disc equation (equation (2) of SF96) as
\begin{equation}
 \f{1}{R}\f{\p}{\p R}\left[\left(\f{3R}{L}\f{\p}{\p R}(\nu_1 L)
 -\f{3}{2}\nu_1 \right){\bf L}+\f{1}{2}\nu_2{RL}\f{\p {\bf l}}{\p R}\right]
 +\f{{\bf \o_p} \times {\bf L}}{R^3}=0,
 \label{pr}
\end{equation}
where $\nu_1$ and $\nu_2$ are the 
kinematic viscosities acting along the plane (horizontal) of the disc 
and normal (vertical) to the disc respectively, and $\o_p/R^3$ is the
Lense-Thirring
precession frequency or the so called orbital plane precession frequency
or the nodal plane precession frequency,
with $\o_p=2GJ/c^2$ ($J$ is the total angular momentum of the black hole,
$G$ is the Gravitational constant and $c$
is the speed of light in vacuum). The modulus of the angular momentum
vector per unit area of the disc can be defined as
\begin{eqnarray}
 L=|{\bf{L}}|=|L{\bf{l}}|=(GMR)^{\f{1}{2}}\Sigma
 \label{Lb}
\end{eqnarray}
where $M$ is the mass of the black hole and $\Sigma$ 
is the surface density of the disc. 

Now, we can take the scalar product of ${\bf l}$ with equation~(\ref{pr}) which turns out to be
\begin{equation}
 \f{1}{R}\f{\p}{\p R}\left[R\f{\p}{\p R}(\nu_1 L)
 -\f{1}{2}\nu_1 L\right]=0 .
 \label{pr1}
\end{equation}
Integrating the above equation (we assume here that viscosity $\nu_1$ is constant) one can obtain 
\begin{equation}
 L=C_2 R^{\f{1}{2}}-2 C_1
 \label{L}
\end{equation}
where $C_1$ and $C_2$ are the integral constants which have to be determined.
As mentioned in Section~\ref{Introduction}, both SF96 and \citet{chen}
discarded the constant $C_1$, which is connected with the inner disc (see Section~\ref{cb}). 
But we keep $C_1$, because we are particularly interested in the 
inner disc in this paper.


\subsection{Discussion on the integral constants $C_1$ and $C_2$}\label{cb}

The expression of $C_2$ is mentioned in SF96, which can be derived by
substituting equation~(\ref{Lb}) into equation~(\ref{L}) and rewriting $C_2$ as
\begin{eqnarray}
 C_2 &=& L R^{-\f{1}{2}}+2C_1 R^{-\f{1}{2}} \nonumber 
 \\ &=& (GM)^{\f{1}{2}}\S+2 C_1 R^{-\f{1}{2}}.
\end{eqnarray}
Now, at $R \rightarrow \infty$ we can substitute
$\S \rightarrow \S_{\infty}$ in the above equation. This gives us 
\begin{eqnarray}
C_2=(GM)^{\f{1}{2}}\S_{\infty}.
 \label{c2}
\end{eqnarray}
Therefore, one can conclude that $C_2$ is connected with the outer disc 
boundary condition. Considering $C_1$, we can write 
\begin{eqnarray}
 2 C_1 &=& (GM)^{\f{1}{2}}R^{\f{1}{2}}\S_{\infty}-L \nonumber
 \\
 &=& (GM)^{\f{1}{2}}R^{\f{1}{2}}[\S_{\infty}-\S (R)].
 \label{c1}
\end{eqnarray}
But, since $C_1$ is a constant,
\begin{equation}
R^{\f{1}{2}}[\S_{\infty}-\S(R)]=\mbox{constant}=-\kappa,
 \label{con}
\end{equation}
where $\kappa$ is considered to be a positive constant. This implies that surface density
increases as $R$ decreases, which is reasonable \citep[e.g., ][]{fkr, lop}. Equation~(\ref{con}) gives
\begin{equation}
\S(R)=\S_{\infty}+\kappa R^{-\f{1}{2}}, 
 \label{lubow}
\end{equation}
which shows that $\S$ varies linearly with $R^{-\f{1}{2}}$.

Considering that the inner edge radius of the accretion disc is $R_{\rm in}$, and 
$\S = \S_{\rm in}$ at $R = R_{\rm in}$, equation~(\ref{c1}) gives
\begin{eqnarray}
	C_1 &=& \frac{1}{2} (GM)^{\f{1}{2}}R_{\rm in}^{\f{1}{2}}[\S_{\infty}-\S_{\rm in}].
 \label{cp}
\end{eqnarray}
This equation shows that $C_1$ is connected with the condition
at the inner boundary, as mentioned (but not shown) by \citet{chen}.
Therefore, one needs to understand the physics of the 
inner disc boundary in order to theoretically estimate the value of $C_1$.
Here we note that the lowest value of $R_{\rm in}$ is the ISCO 
radius $R_{\rm ISCO}$ for a black hole. For the prograde rotation of the disc,
$R_{\rm ISCO}$ has a range of $6R_g$ to $R_g$ for a non-spinning to a maximally
spinning black hole. Here, $R_g~(= GM/c^2)$ is the gravitational radius.

Therefore, two boundary conditions from the inner disc and the outer disc
give the expressions for $C_1$ and $C_2$ (equations~\ref{c2} and \ref{cp}).
From these and equation~(\ref{L}) we get,
\begin{eqnarray}
L=(GM)^{\f{1}{2}}[R^{\f{1}{2}}\S_{\infty}+R_{\rm in}^{\f{1}{2}}(\S_{\rm in}-\S_{\infty})].
\label{Lp}
\end{eqnarray}
Now, using equation~(\ref{con}) and the inner disc boundary condition, 
$\kappa$ can be written as
\begin{eqnarray}
	\kappa=R_{\rm in}^{\f{1}{2}}[\S_{\rm in}-\S_{\infty}],
\end{eqnarray}
and substituting it in equation~(\ref{lubow}) we get
\begin{eqnarray}
 \S  (R)&=&\S_{\infty}+\left(\f{R_{\rm in}}{R}\right)^{\f{1}{2}}[\S_{\rm in}-\S_{\infty}] \nonumber
 \\
 &=& \S_{\infty}-\frac{2C_1}{(GMR)^{\f{1}{2}}}.
 \label{sigfinal}
\end{eqnarray}

Equation (\ref{sigfinal}) shows that, if $C_1$ is zero
\citep[as considered by SF96 and ][]{chen}, then $\S(R) = \S_{\infty}$,
implying that $\S$ is independent of $R$. 
The radial variation of $\S$, i.e., the second term of equation (\ref{sigfinal}),
plays a significant role in the structure of the disc.

\subsection{Lense-Thirring precession in slow-spinning limit}
\label{cc}

In Cartesian coordinates with the black hole centre as the origin and
the $z$ axis along ${\bf \o_p}$, we can write (SF96),
\begin{equation}
 {\bf \o_p} \times {\bf l}=(-\o_p l_y, \o_p l_x, 0).
\end{equation}
Now, substituting equation~(\ref{L}) into equation~(\ref{pr}) and combining the 
$x$ and $y$ components, the warped disc equation can be written as (equation (6) of SF96)
\begin{equation}
\f{\p}{\p R}\left(3\nu_1 C_1 W
 +\f{1}{2}\nu_2{RL}\f{\p W}{\p R}\right)=-i\o_p\f{L}{R^2}W, 
 \label{main}
\end{equation}
where $W=l_x+il_y=\b e^{i\g}$. Here, $\b=|W|$ and $\g$ represent the 
tilt and the twist of the disc respectively as a function of $R$
due to the frame-dragging effect.
SF96 mentioned that equation~(\ref{main}) is at least quite difficult, 
if not impossible, to solve analytically.
Hence, SF96 discarded the term $C_1$ to make equation~(\ref{main})
analytically solvable.
However, it is possible to solve this equation analytically, even keeping $C_1$, if the parameter $W$ is 
expanded around the Kerr parameter $a$, and the series is truncated after the linear term. 
We do such a truncation in this paper.

Here we note that the slow-spinning limit of the Lense-Thirring precession frequency has been used
by many authors \citep[e.g., ][]{bp,chen,nx,p72,p92,sf,zhu}. This limit, which is written as
\begin{eqnarray} 
\O_{nod}  \approx  \f{2GJ}{c^2R^3} = \f{\bf{\o_p}}{R^3} 
= 2 a  \f{G^2M^2}{c^3 R^3},
 \label{ltlow}
\end{eqnarray}
can be obtained from a more general expression \citep[for prograde rotation $(a > 0)$; ][]{kato}
\begin{eqnarray} 
\O_{nod}= \f{(GM)^{\f{1}{2}}} {\left(R^{\f{3}{2}}+aR_g^{\f{3}{2}}\right)}.
 \left[1-\left(1-\f{4aR_g^{\f{3}{2}}}{R^{\f{3}{2}}}
 +\f{3a^2R_g^2}{R^2} \right)^{\f{1}{2}}\right], \nonumber
 \label{lt}
\end{eqnarray}
using the slow-spinning limit $a << 1$. Since, we use this 
slow-spinning limit (equation~\ref{ltlow}) of the Lense-Thirring precession frequency throughout this paper,
it is consistent to use a slow-spinning limit for $W$, as mentioned above.

\subsection{Analytical solution of the tilted/warped disc equation}\label{wds}
In this section, we present an analytical calculation of the 
accretion disc tilt angle expression, which is the main result of this paper.
As we have discussed in the previous section, we consider a
solution of equation~(\ref{main}) upto the linear order in $a$ (in slow-spinning limit):
\begin{equation}
 W(R)=W_0+aW_a,
 \label{wb}
\end{equation}
where $|W_0|$ is the tilt angle for an almost non-spinning ($a \approx 0$)
black hole, and $aW_a$ is the change in $W_0$ due to the presence of $a$. Thus, $|W(R)|$
is the total tilt of the accretion disc for a slow spin of a black hole.

If we consider $a \approx 0$, equation~(\ref{main}) reduces to
\begin{equation}
 \f{\p}{\p R}\left(3\nu_1 C_1 W_0+\f{1}{2}\nu_2{RL}\f{\p W_0}{\p R} \right)=0, 
 \label{a0e}
\end{equation}
Solving equation~(\ref{a0e}) we obtain
\begin{equation}
 W_0=\left[K_2\left(\f{\sqrt{R}}{L}\right)^n+K_1\right],
 \label{w0}
\end{equation}
where $n=6\nu_1/\nu_2$. Now, the two arbitrary constants $K_1$ and $K_2$
have to be determined from suitable boundary conditions.
Suppose, (i) $W_0 \rightarrow W_{\infty}$ at $R\rightarrow \infty$, i.e., at the outer edge of the
disc, which is very large, and (ii) $W_0 \rightarrow W_{0,\rm in}$ at $R \rightarrow R_{\rm in}$, i.e.,
at the disc inner edge. Then $K_1$ and $K_2$ will be
\begin{equation}
 K_1=\f{(W_{0,\rm in}-W_{\infty}z_{\rm in}^n)}{1-z_{\rm in}^n}
\end{equation}
and
\begin{equation}
 K_2=C_2^n\f{(W_{\infty}-W_{0,\rm in})}{1-z_{\rm in}^n}.
\end{equation}
Finally, we can express equation~(\ref{w0}) as
\begin{equation}
 W_0=\f{z^n(W_{\infty}-W_{0,\rm in})+(W_{0,\rm in}-W_{\infty}z_{\rm in}^n)}{1-z_{\rm in}^n},
 \label{iwlt}
\end{equation}
where, 
\begin{eqnarray}
z_{\rm in}&=&1+\f{2C_1}{L_{\rm in}}=\f{\S_{\infty}}{\S_{\rm in}}
 \label{zin}
 \\
 {\rm and} ~~~ z&=&1+\f{2C_1}{L} = \f{\S_{\infty}\sqrt{R}}{\S_{\infty}\sqrt{R}+(\S_{\rm in}
-\S_{\infty})\sqrt{R_{\rm in}}} \nonumber
\\
&=& \f{z_{\rm in}\sqrt{R}}{z_{\rm in}\sqrt{R}+(1-z_{\rm in})\sqrt{R_{\rm in}}}
\label{z}
\end{eqnarray}
are expressed using equations (\ref{cp}) and (\ref{Lp}).
Since we are mainly interested in the tilt of the disc, and not the twist, in this paper,
we assume $\gamma \approx 0$ at the boundaries, and hence $W_{\infty}$ and $W_{0,\rm in}$ are real.
Therefore, $W_0$ in equation~(\ref{iwlt}) is also a real quantity.
equation~(\ref{iwlt}) shows that the tilt angle ($|W_0|$) can be non-zero and a 
function of $R$ depending on the boundary conditions, even though frame-dragging or 
LT precession is almost absent. Thus LT precession is not necessary to tilt an accretion disc. 
Note that \cite{mp} also derived an expression (their equation~(A5)) of
a $R$ dependent tilt angle in the absence of LT precession.

We further note that, if the LT precession vanishes $(\o_p=0)$, the equation~(8) of SF96
(which considers $C_1 = 0$) gives $W=K$, which is same as the tilt at infinity. 
This implies that, according to SF96, the tilt angle is independent of $R$ if there is no LT precession.
Here we show that this conclusion is only a result of the $C_1 = 0$ assumption of SF96.
If we consider $C_1 = 0$ and $a \approx 0$, we get $W_0=W_{s}$
(see equation (\ref{ws1}) of APPENDIX \ref{appendix}) from our calculations. Now with $R_{\rm in} \rightarrow 0$, which is another assumption
made in SF96, we get $W_0=W_{\infty} \equiv K$ of SF96. Thus, 
while the tilt angle generally depends on $R$ even for $a \approx 0$ (equation~\ref{iwlt}), it is a constant
consistent with SF96, if the inner disc is ignored by assuming $C_1 = 0$.

Now substituting equation~(\ref{wb}) into equation~(\ref{main}) and neglecting the
higher order terms $(O(a^2))$, we can obtain 
\begin{equation}
\f{\p}{\p R}\left(3\nu_1 C_1 W_a
 +\f{1}{2}\nu_2{RL}\f{\p W_a}{\p R}\right)=-2iq\f{L}{R^2}W_0, 
 \label{wa}
\end{equation}
where $q=G^2M^2/c^3$ is a constant. 
Note that the term related to $\p W_0/\p R$ disappears from 
equation (\ref{wa}) as $W_0$ satisfies equation (\ref{a0e}).

Integrating equation~(\ref{wa}) we obtain
\begin{eqnarray}
&&3 \nu_1 C_1 W_a+\f{1}{2}\nu_2{RL}\f{\p W_a}{\p R}=K_3 \nonumber
\\
&+& 2iq\f{2C_1(n-2)(W_{0,\rm in}-W_{\infty}z_{\rm in}^n)(L+C_1)+L^2z^n(W_{\infty}
-W_{0,\rm in})}{(1-z_{\rm in}^n)(n-2)C_1R}, \nonumber
\\
\label{dwdr1}
\end{eqnarray}
where $K_3$ is an integral constant. equation~(\ref{dwdr1}) can be written as
\begin{eqnarray}
&&\f{d W_a}{d R}+\f{n C_1}{RL} W_a=\f{2K_3}{\nu_2 RL}
+\f{4iq}{(1-z_{\rm in}^n)(n-2)C_1\nu_2R^2L}. \nonumber
\\
&&[2C_1(n-2)(W_{0,\rm in}-W_{\infty}z_{\rm in}^n)(L+C_1)+L^2z^n(W_{\infty}-W_{0,\rm in})]. \nonumber
 \label{dwdr}
 \\
\end{eqnarray}
 The exact solution of equation~(\ref{dwdr}) is
\begin{eqnarray}
 W_a&=&\f{1}{2n(n+2)\nu_2C_1^2R}.\left[i\nu_2uC_2^2R+i4nC_1^3(n+2)v\nu_2z^n \right. \nonumber
 \\
&& \left. -i2n\nu_2C_1^2(u+2C_2(n+2)vR^{\f{1}{2}}z^n) \nonumber
 \right.\\ 
&& \left.+2C_1(2K_3(n+2)R+inC_2\nu_2uR^{\f{1}{2}})\right]+K_4\left(\f{\sqrt{R}}{L}\right)^n, \nonumber
\\
 \label{wp}
\end{eqnarray}
where,
\begin{eqnarray}
 u&=&\f{8q(W_{0,\rm in}-W_{\infty}z_{\rm in}^n)}{\nu_2(1-z_{\rm in}^n)}
 \\
 {\rm and} ~~~  v&=&\f{4q(W_{\infty}-W_{0,\rm in})}{\nu_2C_1(n-2)(1-z_{\rm in}^n)}.
\end{eqnarray}

Since the frame-dragging should not affect the tilt of the disc far away from the
central black hole, we assume $W_a \rightarrow 0$ for $R\rightarrow \infty$.
Therefore, at $R\rightarrow \infty$, $W = W_0 \rightarrow W_{\infty}$.
Using this boundary condition, $K_4$ can be determined as
\begin{eqnarray}
 K_4=-\f{C_2^n}{2n(n+2)\nu_2C_1^2}\left[iu\nu_2C_2^2+4C_1(n+2)K_3\right] .
 \label{k4}
\end{eqnarray}
Substituting this expression of $K_4$ in equation~(\ref{wp}), 
and assuming $W_a \rightarrow W_{a,\rm in}$ at $R\rightarrow R_{\rm in}$
($W_{a,\rm in}$ is considered real, like $W_{\infty}$ and $W_{0,\rm in}$),
we obtain $K_3$ as
\begin{eqnarray}
&& \f{2K_3}{nC_1\nu_2}=\f{W_{a,\rm in}}{1-z_{\rm in}^n} \nonumber
 \\
&-& \f{i\left[uC_2^2(1-z_{\rm in}^n)+\f{2nC_1(L_{\rm in}+C_1)}{R_{\rm in}}
[u-2C_1(n+2)vz_{\rm in}^n]\right]}{2n(n+2)C_1^2(1-z_{\rm in}^n)}. \nonumber
\\
 \label{k3}
\end{eqnarray}
Substituting $K_3$ and $K_4$ in equation~(\ref{wp}), we obtain the final
expression of $W_a$ as
\begin{eqnarray}
 W_a &=& \f{W_{a,\rm in}(1-z^n)}{1-z_{\rm in}^n}
 +\f{8iq}{\nu_2 C_1(n^2-4)(1-z_{\rm in}^n)}. \nonumber
 \\
&& \left[\left(\f{L+C_1}{R}\right).
 [(n-2)(W_{0,\rm in}-W_{\infty}z_{\rm in}^n) \right. \nonumber
 \\
&& \left.-(n+2)(W_{\infty}-W_{0,\rm in})z^n]
-\left(\f{1-z^n}{1-z_{\rm in}^n}.\f{L_{\rm in}+C_1}{R_{\rm in}}\right). \right.  \nonumber
 \\
&& \left. [W_{0,\rm in}((n-2)+(n+2)z_{\rm in}^n)-2nW_{\infty}z_{\rm in}^n] ~\right]\nonumber .
\\
\label{Waf}
\end{eqnarray}
Now, using equations (\ref{cp}) and (\ref{Lp}) we replace $L, ~L_{\rm in}$ and $C_1$ from
equation (\ref{Waf}) and using
equations (\ref{iwlt}) and (\ref{Waf}) we can express equation~(\ref{wb}) as
\begin{eqnarray}
 W=W_0+a~W_a = A+iB,
\end{eqnarray}
where, $A$ and $B$ can be expressed as 
\begin{eqnarray}
 A=\f{1}{1-z_{\rm in}^n}.\left[W_{\infty}(z^n-z_{\rm in}^n)+(1-z^n)(W_{0,\rm in}+aW_{a,\rm in})\right]
 \label{A}
\end{eqnarray}
and
\begin{eqnarray}
 B &=& \f{8aq}{\nu_2 (n^2-4)(1-z_{\rm in}^n)}.\left[-\f{2z_{\rm in}\sqrt{R}
 +\sqrt{R_{\rm in}}(1-z_{\rm in})}{R(1-z_{\rm in})\sqrt{R_{\rm in}}}. \right. \nonumber
\\
 && \left. [(n-2)(W_{0,\rm in}-W_{\infty}z_{\rm in}^n)-(n+2)(W_{\infty}-W_{0,\rm in})z^n] \right. \nonumber
\\
&&\left. +\left(\f{1-z^n}{1-z_{\rm in}^n}.\f{1+z_{\rm in}}{R_{\rm in}(1-z_{\rm in})}\right). \right. \nonumber
\\
&&\left. [W_{0,\rm in}((n-2)+(n+2)z_{\rm in}^n)-2nW_{\infty}z_{\rm in}^n]~ \right]. \nonumber
\label{B}
\\
\end{eqnarray}
Here, $q=G^2M^2/c^3$ and $n=6\nu_1/\nu_2$. The expressions of $z_{\rm in}$ and $z$ are
given in equations (\ref{zin}) and (\ref{z}) respectively. Finally, the 
tilt angle can be expressed as
\begin{eqnarray}
 |W|=\sqrt{A^2+B^2} .
 \label{mW}
\end{eqnarray}
Therefore, in our formalism, the tilt angle at a given radial distance $R$ depends on 
the following {\it free} parameters:
$M$, $a$, $\nu_1$, $\nu_2$, $R_{\rm in}$, $z_{\rm in}(=\S_{\infty}/\S_{\rm in})$,
$W_{\rm 0, in}$, $W_{\rm a, in}$ and $W_{\infty}$.

\subsection{Upper limit of $a$}\label{rangea}

Equation (\ref{wb}) can be valid if 
\begin{eqnarray}
 |W_0| > a |W_a|.
 \label{a}
\end{eqnarray}
Using equations (\ref{iwlt}) and (\ref{Waf}),
$|W_0|$ and $|W_a|$ can be written as
\begin{eqnarray}
 |W_0|=\f{1}{1-z_{\rm in}^n}.\left[W_{\infty}(z^n-z_{\rm in}^n)+(1-z^n)W_{0,\rm in}\right]
 \label{W_0}
\end{eqnarray}
and
\begin{eqnarray}
 |W_a|^2 &=&\left[\f{W_{a,\rm in}(1-z^n)}{1-z_{\rm in}^n}\right]^2+\f{64q^2}{[\nu_2 (n^2-4)(1-z_{\rm in}^n)]^2}. \nonumber
 \\
&&\left[-\f{2z_{\rm in}\sqrt{R}
 +\sqrt{R_{\rm in}}(1-z_{\rm in})}{R(1-z_{\rm in})\sqrt{R_{\rm in}}}. \right. \nonumber
\\
 && \left. [(n-2)(W_{0,\rm in}-W_{\infty}z_{\rm in}^n)-(n+2)(W_{\infty}-W_{0,\rm in})z^n] \right. \nonumber
\\
&&\left. +\left(\f{1-z^n}{1-z_{\rm in}^n}.\f{1+z_{\rm in}}{R_{\rm in}(1-z_{\rm in})}\right). \right. \nonumber
\\
&&\left. [W_{0,\rm in}((n-2)+(n+2)z_{\rm in}^n)-2nW_{\infty}z_{\rm in}^n]~ \right]^2. 
\label{W_a}
\end{eqnarray}
Using equations~(\ref{W_0}) and (\ref{W_a}) with assumed parameter values, 
an upper limit ($a_{\rm u}$) of $a$ can be determined in such a way 
that equation~(\ref{a}) is satisfied for all $R$ values, i.e., for the entire disc. Our results are
valid when $a < a_{\rm u}$.

\section{Behaviour of the tilt angle}\label{results}

\subsection{Parameter values}\label{Parameter}

In order to explore the nature of the tilt angle radial profile, we need to 
choose suitable values of the free parameters appeared in the expressions of $A$ and $B$
(equations~(\ref{A}) and (\ref{B})), as mentioned in the end of Section~\ref{wds}.
Since the measured mass values of Galactic accreting black holes are consistent with 
$\sim 5-15 M_\odot$ \citep{Fragos2015}, we consider $M=10M_{\odot}$ for the purpose of
demonstration. For our slow-spinning limit and prograde rotation, $R_{\rm in}$ is very close to, but less than,
$6R_g$ (see Section~\ref{cb}). Therefore, we use $R_{\rm in}=6R_g$ for all cases.
Considering this $R_{\rm in}$ value and for a given set of parameter values, we calculate 
$a_{\rm u}$ (Section~\ref{rangea}), and use values of $a$ less than this $a_{\rm u}$.

The disc surface density $\S_{\infty}$ far away from the black hole 
could be estimated from an inferred accretion rate $(\dot{M})$
by the relation $\dot{M}=3\pi \nu_1 \S_{\infty}$ \citep[e.g., ][]{natp}.
But, we need only the ratio between two surface density values, i.e.,
$\S_{\infty}/\S_{\rm in}$ ($= z_{\rm in}$; see equations~(\ref{A}) and (\ref{B})).
As this ratio is less than $1$ in our formalism (Section~\ref{cb}), we assume 
$\S_{\infty}/\S_{\rm in} = 0.75$ for the purpose of demonstration.
The viscosity of an accretion disc could be of the order of $10^{14}$ cm$^2$ s$^{-1}$ 
\citep[e.g., ][]{fkr}, and we consider a range of $10^{14} - 10^{15}$ cm$^2$ s$^{-1}$.
We use a range $0.5 - 1$ for the viscous anisotropy $\nu_2/\nu_1$, which is consistent
with the results of \citet{mp} (see Section~\ref{Introduction}).
We consider modest values ($0^\circ - 10^\circ$) of the tilt angle $W_{\rm in}$ 
at the disc inner edge. This is not only consistent with some
previous theoretical computations \citep[e.g., ][]{iv, lop, zhul},
but also similar to an observationally inferred tilt angle
\citep[section 6.3 of ][]{Ingrametal2017}.
We use $W_{a,\rm in}=0^\circ$, implying $W_{0,\rm in} = W_{\rm in}$, and $W_{\infty}=10^\circ$ 
\citep[e.g., ][]{p71} for the purpose of demonstration.

\begin{figure}
  \begin{center}
\includegraphics[width=3.5in,angle=0]{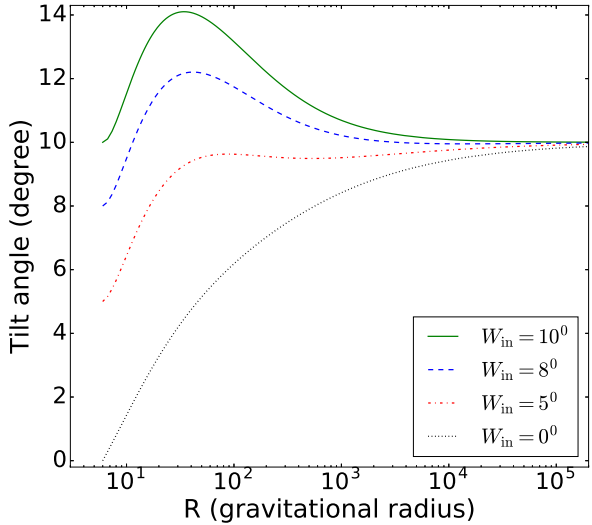}
\caption{\label{fg1} Radial profiles of the accretion disc tilt angle using equation~(\ref{mW})
for four different values of the disc inner edge tilt 
angle $W_{\rm in}$ (including $W_{\rm in}=0^\circ$). 
The other parameter values are $M = 10M_{\odot}$, $a=0.007$, $\nu_1 = 
\nu_2 = 10^{14}$ cm$^2$ s$^{-1}$ and $\S_{\infty}/\S_{\rm in}=0.75$. 
This figure shows that the inner part of the disc is tilted for all values of
$W_{\rm in}$ (even for $W_{\rm in}=0^\circ$).
\label{fig1}}
\end{center}
\end{figure}

\begin{figure}
  \begin{center}
 \includegraphics[width=3.5in,angle=0]{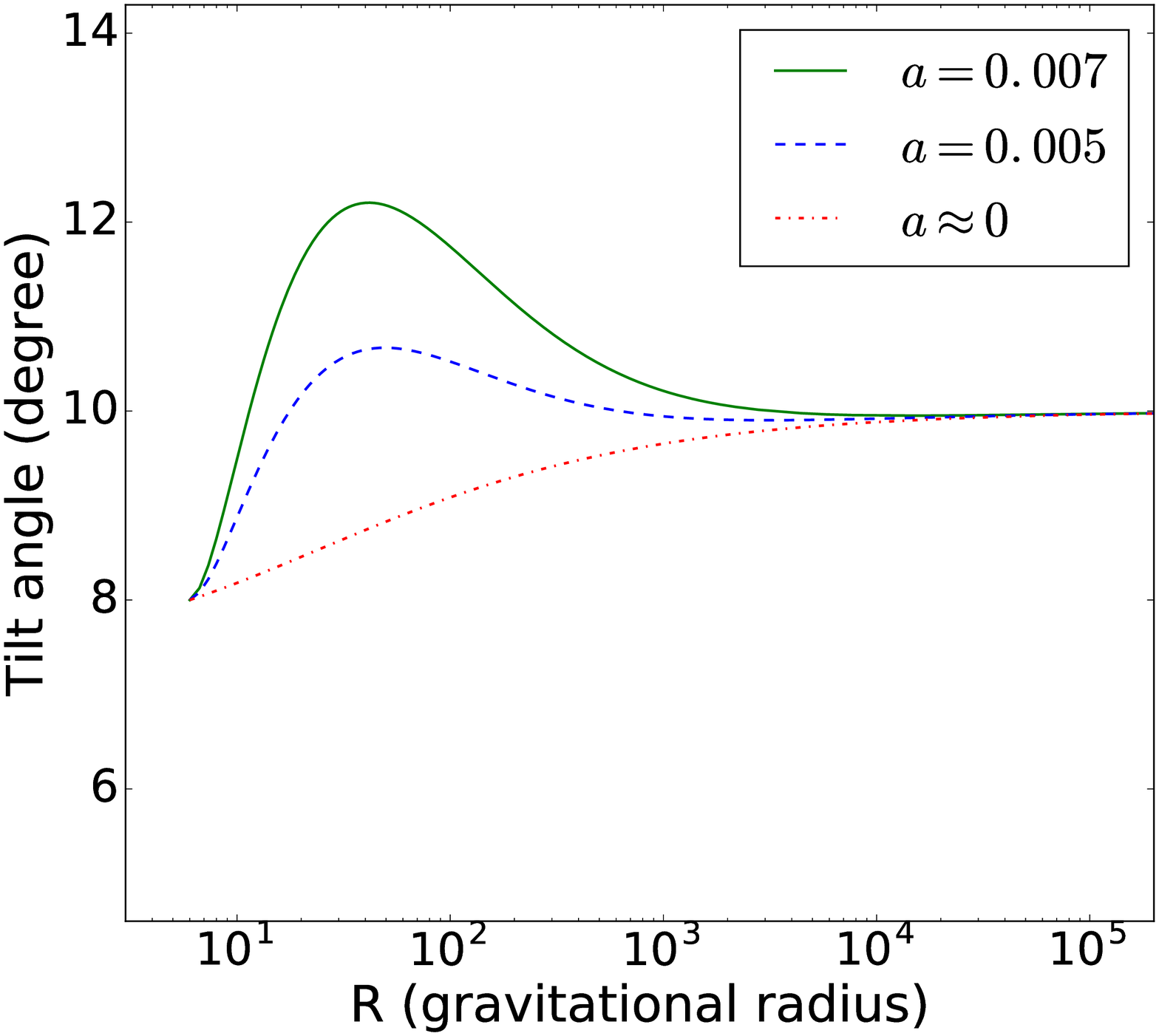}
\caption{\label{fg2} Radial profiles of the tilt angle using equation~(\ref{mW}) 
for three values of the Kerr parameter $a$. Other parameter values are same as
those for the blue dashed curve of Figure~\ref{fg1}. 
\label{fig2}}
\end{center}
\end{figure}

\begin{figure}
  \begin{center}
 \includegraphics[width=3.5in,angle=0]{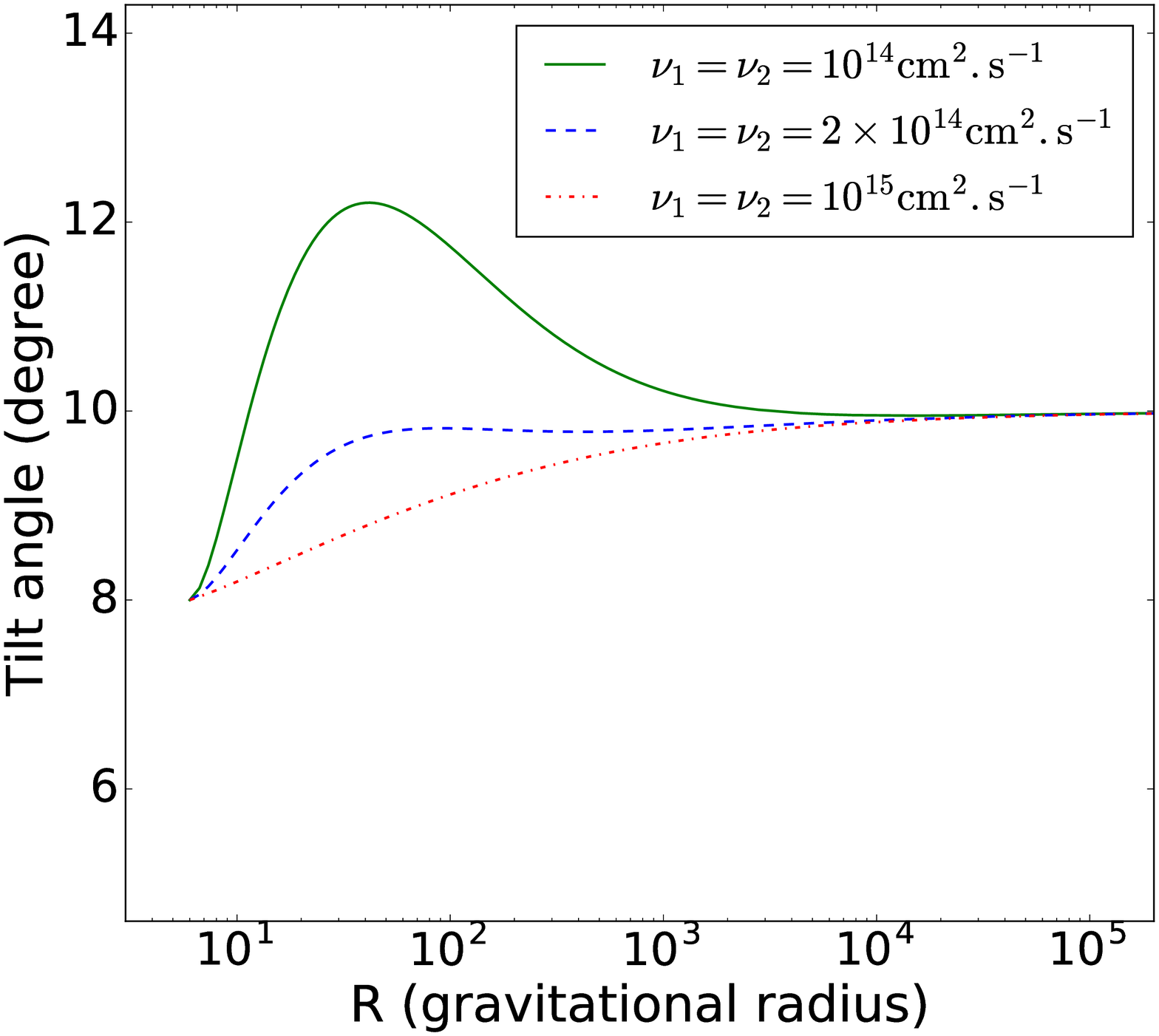}
 \caption{\label{fg3} Radial profiles of the tilt angle using equation~(\ref{mW})
 for three values of the viscosity $\nu_1 = \nu_2$. Other parameter values are same as
 those for the blue dashed curve of Figure~\ref{fg1}. 
\label{fig3}}
\end{center}
\end{figure}

\begin{figure}
  \begin{center}
\includegraphics[width=3.5in,angle=0]{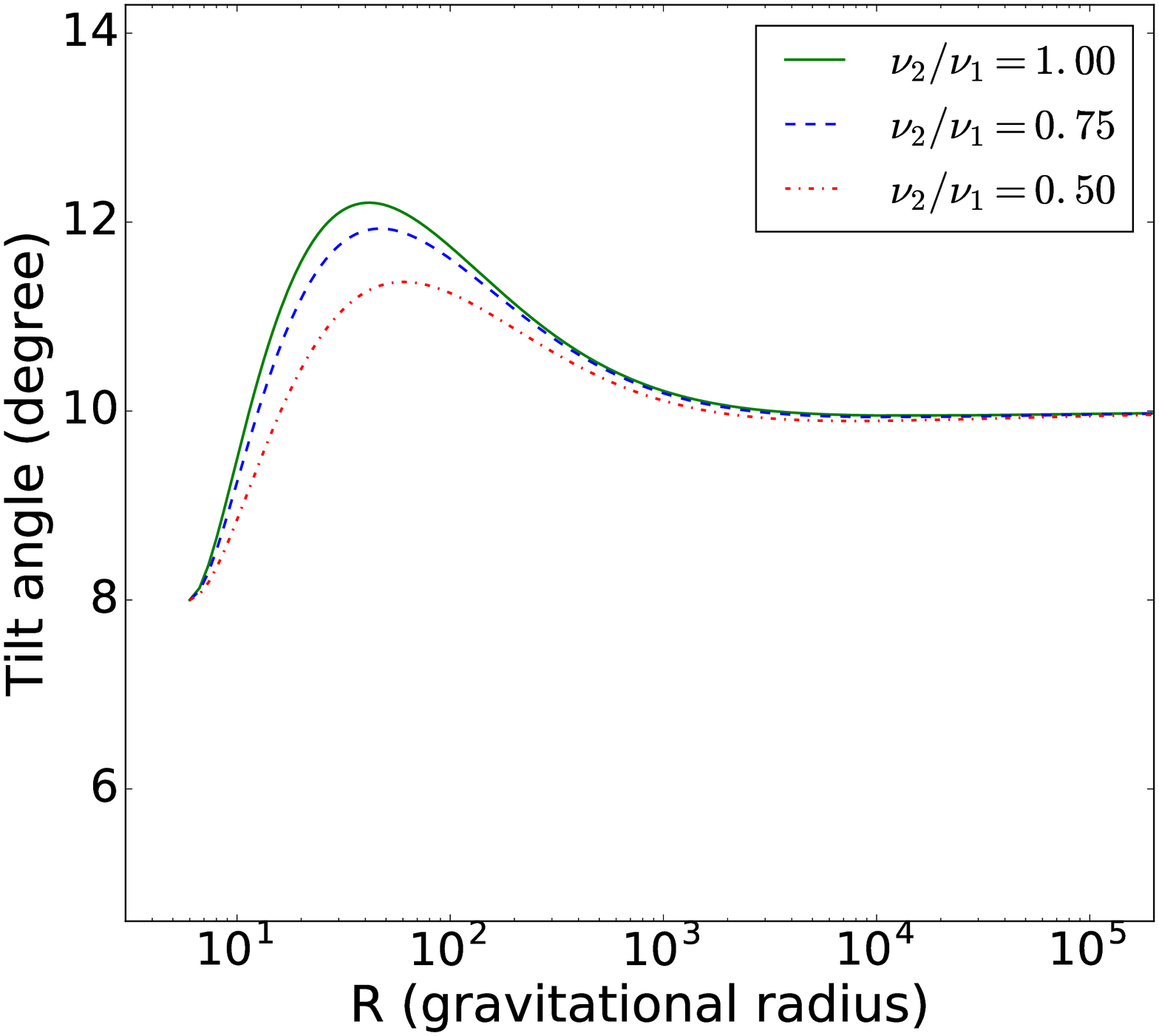}
\caption{\label{fg4} Radial profiles of the tilt angle using equation~(\ref{mW})
	for three values of the viscous anisotropy $\nu_2/\nu_1$. Other parameter values are same 
	(except $\nu_1$) as those for the blue dashed curve of Figure~\ref{fg1}. 
 \label{fig4}}
\end{center}
\end{figure}

\subsection{Discussion using radial profiles of the tilt angle}\label{bw}

We present radial profiles of the tilt angle $|W|$ (equation~\ref{mW}) for 
various example sets of {\it free} parameter values in this section. 
Here, we specifically study the effects of the disc inner edge tilt angle $W_{\rm in}$, 
the Kerr parameter $a$, and the viscosities $\nu_1$ and $\nu_2$ on the
nature of the $|W|$ profile.

Note that our disc inner edge boundary condition is $W = W_{\rm in}$,
where $W_{\rm in}$ is a {\it free} parameter. This means we do not make
any specific assumption about this boundary condition.
For example, we do not assume a zero torque at $R=R_{\rm in}$,
which, although often used \citep[but not always; see, for example, ][]{balbus, pen, kul, zhu12},
is an ad hoc assumption \citep[e.g., ][]{mc14}. Moreover, a zero torque at $R=R_{\rm in}$
is unlikely because of the frame-dragging effect. Here, we explore $|W|$ profiles for various 
$W_{\rm in}$ values (both zero and non-zero; see Figure~\ref{fg1}), but do not attempt to constrain $W_{\rm in}$
in this paper. Such constraint could come from a theoretical understanding of the 
inner disc boundary physics,
but more likely from observations \citep[e.g., see section 6.3 of ][]{Ingrametal2017}.
Our solution with $W_{\rm in}$ as a {\it free} parameter is ideal to confront such observations
with the theory.
	
Figure~\ref{fg1} shows four $|W|$ profiles for different $W_{\rm in}$ values, but 
for same $W_{\infty}$ and other parameter values. For all the $W_{\rm in}$ values,
even for $W_{\rm in}=0$, the tilt angle in the inner part of the disc (e.g., $10R_g-100R_g$)
is at least a significant fraction of $W_{\infty} (=10^\circ)$. This is somewhat different 
from the $\beta$ (same as $|W|$) curve of Figure 3 of \cite{bp}, and suggests
a lesser degree of alignment of the inner disc in our case compared to that suggested by \cite{bp}, 
even when we set the $W_{\rm in}=0$ boundary condition.

Another important aspect of $|W|$ profiles is the tilt angle does
not decrease monotonically to a lower value at the disc inner edge
in some cases (Figure~\ref{fg1}).
Rather, $|W|$ increases with the decrease of $R$ in the inner disc 
for some particular parameter values, and then decreases to the inner edge boundary value.
For example, the green solid curve ($W_{\infty} = W_{\rm in} = 10^\circ$) of 
Figure~\ref{fg1} shows that the tilt angle becomes significantly greater than 
$10^\circ$ in the inner disc, before coming down to $10^\circ$ at the disc inner edge.
This is an effect of the non-zero Kerr parameter value, as the tilt angle would remain $10^\circ$
throughout the disc for $a \approx 0$, which can be easily verified from equation~(\ref{iwlt})
using $W_{0,\rm in} = W_{\infty}$.
Here we note that the increase of the tilt angle with decreasing radial distance
in the inner disc was reported earlier by several authors 
\citep{iv,lop,fra}, but using different methods. Among these, \citet{lop} considered the 
presence of a steady wave-like shape of a prograde warped disc, which is similar to 
the humps shown in Figure~\ref{fg1}. It is interesting to note that the corresponding
radial wavelength goes as $a^{-1/2}$, which implies the disappearance of the hump
for $a \approx 0$, like in our case.

In Figure~\ref{fg2}, we display the tilt angle profiles for three Kerr parameter values: $a = 0.007$,
$a = 0.005$ and $a \approx 0$, keeping other parameter values same. These curves show that,
for a higher $a$ value, the tilt angle is larger and 
the hump of the $|W|$ profile is more pronounced. In Figure~\ref{fg2}, we use
$W_{\rm in}=8^{\circ}$ for all the curves for the purpose of demonstration.
But even for $W_{\rm in}=0$, we find that a higher $a$ value corresponds to a larger tilt angle.
This is expected, because a larger $a$ value, via a stronger LT effect, 
should align the inner disc with the black hole spin equatorial plane 
more sharply (Section~\ref{Introduction}),
and that is possible for fixed values of $W_{\rm in}$ and $W_{\infty}$
only if a higher $a$ value corresponds to a larger tilt angle.

Figure~\ref{fg3} displays the tilt angle profiles for three viscosity values, and shows that
the effect of viscosity is opposite to that of the Kerr parameter. This is 
expected as viscosity opposes the LT effect (Section~\ref{Introduction}).
Figure~\ref{fg4} shows that, if the viscous anisotropy $\nu_2/\nu_1$ value decreases
from $1$ to $0.5$ keeping other parameter values same, 
tilt angle at the inner disc also decreases. This means that the value of the tilt angle 
is lower for a higher horizontal viscosity $(\nu_1)$ with a fixed  vertical
viscosity $(\nu_2)$.

\section{Summary and Implications}\label{conclu}

In this paper, we analytically solve a warped accretion disc equation 
considering a prograde disc around a slowly spinning black hole.
In our exact solution, we do not neglect the inner disc, 
and extend our disc up to a very large distance.
Therefore, our solution should be valid for the inner disc as well
as for the outer disc.

The inner accretion disc can act as a rare natural 
probe to understand the elusive fundamental physics of the strong gravity regime. 
The spectral and timing properties of the X-ray emission from such a disc 
make the theoretical predictions regarding the strong gravity regime testable.
These properties significantly depend on the tilt of the inner disc 
with respect to the black hole spin equator.
Therefore, it can be very useful to have the disc inner edge tilt angle $W_{\rm in}$
as a {\it free} parameter in the solution of a disc equation, 
because $W_{\rm in}$ could be observationally
inferred \citep[e.g., see section 6.3 of ][]{Ingrametal2017}.
Our analytical solution has $W_{\rm in}$ as a free parameter.
Moreover, some other free parameters (e.g., $M$, $a$)
included in our solution can also be observationally estimated
\citep[e.g., ][]{Fragos2015,Miller2007}.
Therefore, our solution for the radial profile of accretion disc tilt angle
can be ideal to confront observations. 
Finally, the recent report of the Fe line energy variation with the phase
of a QPO from a black hole, possibly due to a tilt of the inner disc (Section~\ref{Introduction}),
has made this work timely. More such phenomena could be discovered and studied 
using current and future space missions, such as {\it AstroSat} and {\it Athena}.

\section*{Acknowledgements}

SB thanks A. Ingram for a useful discussion. We thank an anonymous
referee for constructive comments that helped to improve the manuscript.

\appendix
\section{}
\label{appendix}
If we consider $C_1=0$ and $a=0$, equation~(\ref{main}) reduces to 
\begin{equation}
\f{d}{dR}\left(R^{\f{3}{2}}\f{\p W_s}{\p R}\right)=0.
 \label{ap1}
\end{equation}
The solution of equation~(\ref{ap1}) is 
\begin{eqnarray}
 W_{s}= -\f{2K_{s1}}{\sqrt{R}}+K_{s2}.
 \label{ws0}
\end{eqnarray}
Using two boundary conditions: (i) $W_{s} \rightarrow W_{\infty}$ at $R \rightarrow \infty$, 
and (ii) $W_{s} \rightarrow W_{\rm in}$ at $R \rightarrow R_{\rm in}$, we obtain
\begin{eqnarray}
 K_{s2}=W_{\infty}
\end{eqnarray}
and
\begin{eqnarray}
 -2K_{s1}=\sqrt{R_{\rm in}}(W_{\rm in}-W_{\infty}).
\end{eqnarray}
Substituting the values of $K_{s1}$ and $K_{s2}$ in equation~(\ref{ws0}) we obtain
\begin{eqnarray}
 W_{s}= W_{\infty}+\sqrt{\f{R_{\rm in}}{R}}~(W_{\rm in}-W_{\infty}).
 \label{ws1}
\end{eqnarray}

\label{lastpage}

\end{document}